\documentclass[aps,pra,superscriptaddress,amsmath,amssymb,preprintnumbers,floatfix,showpacs,12pt]{revtex4}

\usepackage{amssymb}
\usepackage{epsfig}

\begin{document}
\title{Single mode quantum properties of the codirectional Kerr
nonlinear coupler:
 frequency mismatch and exact solution }
\author{ Faisal A. A. El-Orany}
\affiliation{Department of Mathematics  and Computer Science,
Faculty of Science, Suez Canal University, Ismailia, Egypt}
\author{ M. Sebawe Abdalla}
\affiliation{Mathematics Department, College of Science, King Saud
University, P.O. Box 2455, Riyadh 11451, Saudi Arabia}

\author{ and J. Pe\v {r}ina}
\affiliation{Department of Optics and Joint Laboratory of Optics,
Palack\'{y} University, 17. listopadu 50,
 772 07 Olomouc, Czech Republic}

\date{\today}

\begin{abstract}
In this paper, we investigate the single mode quantum properties
of the codirectional Kerr nonlinear coupler when the frequency
mismatch is involved and a condition for an exact solution of
equations of motion is fulfilled. Particularly, we investigate
quadrature and principal squeezing, Wigner function, quadrature
distribution, phase distribution and phase variance. We show that
the quadrature squeezing and the phase variance can exhibit
collapse-revival and collapse-revival-subrevival phenomena,
respectively, based on the values of the detuning parameter.
Furthermore, we analytically demonstrate that the system can
generate cat states, in particular, Yurke-Stoler states.
\end{abstract}

 \maketitle


\section{Introduction}
Recently, there has been a great interest in the possibility of
using optical devices for ultra-ligh-speed data processing. This
is the main object in the quantum information theory, which aims
at storing and transferring data \cite{EKer1}. One of the
promising devices for data transmission is the nonlinear
directional coupler that consists of two or more parallel optical
waveguides fabricated from some nonlinear material. Both
waveguides are placed close enough to permit flux-dependent
transfer of energy between them by means of evanescent waves. This
flux transfer can be controlled by the device design and the
intensity of the input flux as well. The outgoing fields from the
coupler can be examined as single or compound modes by means of
homodyne detection to observe squeezing of vacuum fluctuations, or
by means of a set of photodetectors to measure photon
correlations, photon antibunching and sub-Poissonian photon
statistics in the standard ways. The investigation of the quantum
properties of light propagating in the directional couplers have
attracted much attention for generating nonclassical light (see
the review paper \cite{qu20} and the references therein).
Moreover, directional coupler is experimentally implemented, e.g.
in planar structures \cite{exp1}, dual optical fibres \cite{exp2}
and certain organic polymers \cite{exp3}.

Among the different types of directional couplers the directional
Kerr nonlinear coupler (DKNC) has taken much attention as a result
of its application in optics as an intensity-dependent routing
switch \cite{jen1,qu10}. DKNC is useful for low intensity fields
and periodical exchange of energy between the guides; but for high
intensity fields, energy is trapped by nonlinearity in the guide
into which it was initially launched (self-trapping effect). The
quantum properties of the DKNC have been studied by several
authors \cite{hora1,hora2,qu14,qu15,qu18,faisal1,ar1}. For
instance, in \cite{qu14} quadrature squeezing, principal squeezing
and integrated intensity variances have been calculated in an
analytical way and investigated based on the transmission of light
between waveguides. Moreover, quantum statistics for
contradirectional KNC have been investigated \cite{ar1} showing
that the switching between waveguides is accompanied by
nonclassical effects, e.g. two-mode squeezing, the generation of
pure state and single mode photon antibunching. Influence of the
geometry of the coupler (when the linear coupling coefficient is
variable) on the statistics of the DKNC \cite{qu15,ar2} (also
contradirectional KNC \cite{ar2}) was investigated showing that
there is a possibility to control energy switching between
waveguides by adjusting the form of a coupling function. The
numerical technique based on the diagonalization of the
Hamiltonian is used for studying the mean-photon number
\cite{qu10} and the phase distribution in the framework of a
quasiprobability distribution function \cite{qu18}. Furthermore,
in the most of papers dealing with DKNC in the literature, e.g.
\cite{qu14,qu15,qu18,faisal1,ar1}, authors have shown that the
evolution of the mean-photon numbers exhibit collapse-revival
phenomenon arising from the nonlinear exchange of energy between
waveguides. In this regard there is a similarity between DKNC and
the behaviour of the atomic inversion in the Jaynes-Cummings model
\cite{jay1}. In the present paper we provide--under a certain
condition--the exact solutions for the equations of motion of DKNC
taking into account the frequency mismatch. This would lead to
many important results additionally to what are already known for
DKNC. For instance, we show that  the quadrature squeezing
 can exhibit collapse-revival phenomenon rather than the mean-photon numbers, as
  shown earlier \cite{qu10, qu14,qu15,faisal1,ar1}.
 Also the phase variances  exhibit collapse-revival-subrevival phenomenon
similar to that of the two-mode Jaynes-Cummings model
\cite{comin}. Moreover, for the first time, we analytically and
numerically prove that the system can generate Yurke-Stoler states
at certain times despite the coupling of the two modes.

Now the Hamiltonian controlling the codirectional Kerr nonlinear
coupler is given as
\begin{equation}
\frac{\hat{H}}{\hbar }=\sum_{j=1}^{2}
[\omega _{j}\hat{a}_{j}^{\dagger }%
\hat{a}_{j}+\chi \hat{a}_{j}^{\dagger 2}
\hat{a}_{j}^{2}]+\widetilde{\chi
}\hat{a}_{1}^{\dagger }\hat{a}_{1}\hat{a}_{2}^{\dagger
}\hat{a}_{2}+\kappa (%
\hat{a}_{1}^{\dagger }\hat{a}_{2}+\hat{a}_{2}^{\dagger }\hat{a}_{1}),
\label{2}
\end{equation}
where $\omega_1$ and $\omega_2$ are the frequencies of the first
and the second modes  with the annihilation operators $\hat{a}_1$
and $\hat{a}_2$, respectively, $\chi$ and $\widetilde{\chi}$ are
the coupling constants proportional to the third-order
susceptibility $\chi^{(3)}$ and responsible for the self-action
and cross-action processes, respectively, $\kappa$ is the linear
coupling constant between the waveguides. Kerr coupler can be
implemented from certain organic polymers with high third-order
nonlinearities \cite{exp3}. Also cento-symmetric optical fibres
can be adopted. We proceed that in above mentioned papers related
to DKNC, i.e. \cite{qu14,qu15,qu18,faisal1,ar1}, the authors have
neglected the non-relevant terms, i.e. the nonlinear rotational
terms, for obtaining a closed form solution of the equations of
motion related to (\ref{2}). In this paper we restrict ourselves
to the case in which the equations of motion can be solved
exactly. In this case the system is linear in the sense that the
mean-photon numbers exhibit oscillatory behaviour indicating
periodic energy exchange between the waveguides even though there
is a nonlinear medium between the waveguides. Actually, this does
not mean that the coupler cannot generate nonclassical effects.
More illustratively,
 the mean-photon numbers are given by the
diagonal elements of the density matrix only.
Nevertheless, the quantities, which depend
on the off-diagonal elements, such as quadratures squeezing, Wigner
function, phase distribution, etc., may generate nonclassical effects.
 Thus we are going to investigate the behaviour of these quantities
in the present paper, considering also the influence of frequency mismatch
$\Delta=\omega_{1}-\omega_{2}$ .
It is worth remembering that the phase mismatches were considered
to the other
types of coupler, e.g. linear coupler \cite{perinova}, nonlinear
coupler
\cite{perj1} and Raman-Brillouin couplers \cite{fura}.

We conclude this section by writing the equations of motion
for (\ref{2}) under the condition $\widetilde{\chi}=2\chi$,
\begin{eqnarray}
\frac{d\hat{a}_{1}}{dt} &=&-i\omega _{1}\hat{a}_{1}-2i\chi (\hat{a}%
_{1}^{\dagger }\hat{a}_{1}+\hat{a}_{2}^{\dagger }\hat{a}_{2})\hat{a}%
_{1}-i\kappa \hat{a}_{2},  \nonumber \\
\frac{d\hat{a}_{2}}{dt} &=&-i\omega _{2}\hat{a}_{2}-2i\chi (\hat{a}%
_{1}^{\dagger }\hat{a}_{1}+\hat{a}_{2}^{\dagger }\hat{a}_{2})\hat{a}%
_{2}-i\kappa \hat{a}_{1}.  \label{3}
\end{eqnarray}
One can easily check that $\hat{a}_{1}^{\dagger
}\hat{a}_{1}+\hat{a}_{2}^{\dagger }%
\hat{a}_{2}=\hat{C}$ is a constant of motion and hence
 the general solution for (\ref{3}) is
\begin{eqnarray}
\begin{array}{lr} \hat{a}_{1}(t)=\exp
(-i\hat{\Lambda}t/2)\Bigl\{
\hat{%
a}_{1}(0)\left[ \cos (\lambda t)-i\frac{\Delta }{2\lambda }\sin
(\lambda t) \right] -i\frac{\kappa }{\lambda }\hat{a}_{2}(0)\sin
(\lambda t)\Bigr\},\\
 \\
\hat{a}_{2}(t) =\exp (-i\hat{\Lambda}t/2)\Bigl\{
\hat{%
a}_{2}(0)\left[ \cos (\lambda t)-i\frac{\Delta }{2\lambda }\sin
(\lambda t)\right] -i\frac{\kappa }{\lambda }\hat{a}_{1}(0)\sin
(\lambda t)\Bigr\},
  \label{4}
\end{array}
\end{eqnarray}
where $\lambda =\sqrt{\kappa ^{2}+\frac{1}{4}\Delta^{2}}$,
$\hat{\Lambda}=\omega_1+\omega _2+4\chi \hat{C}$ and $\Delta$ is
the frequency mismatch. One can easily check that
$\hat{a}_{1}(t)\leftrightarrow \hat{a}_{2}(t)$ when
$\hat{a}_{1}(0)\leftrightarrow \hat{a}_{2}(0)$. Thus we restrict
the discussion to the behaviour of the first mode only.
 It is worth mentioning that for solving the
problem in the space domain we have to use the substitution
$z=t\vartheta$, where $\vartheta$ is the velocity of light in the
waveguide and $z$ is the travelled distance.  Moreover, we do not
consider the dissipation, which generally leads to decrease of the
total number of photons and to the tendency to reduce the exchange
of photons between the waveguides, i.e. to the reduction of the
nonclassical effects \cite{hora1,hora2}.

Finally, using (\ref{4}) we investigate only the single mode
quantum properties for the DKNC. We perform such investigation in
the following order: In section 2 we examine the quadratures and
principal squeezing. In section 3 we calculate and discuss the
quasiprobability distribution and quadrature distribution. In
section 4 we study the phase distribution and its variance. The
results are summarized in section 5.

\section{Quadratures and principal squeezing}
The photons produced in a nonlinear optical device such as DKNC
are known to have unusual correlation properties, which results in
many nonclassical aspects of the radiation field. Thus in this
section we demonstrate the single mode quadratures and principal
squeezing. As is well known squeezed light has less noise than
coherent light in one of the field quadratures. This light can be
measured by homodyne detection where the signal is superimposed on
a strong coherent beam of the local oscillator. Moreover,
squeezing is one of the most important phenomenon in quantum
optics because  of its applications in various areas, e.g., in
optics communication, quantum information theory, etc.\cite{inf1}.

To investigate the single mode squeezing we define two quadratures
$\hat{X}$ and $\hat{Y}$, which denote the real (electric) and
imaginary (magnetic) parts, respectively, of the radiation field,
as
\begin{equation}
 \hat{X}=\frac{1}{\sqrt{2}}[\hat{A}_{1}(t)+
\hat{A}^{\dagger}_{1}(t)],\quad
 \hat{Y}=\frac{1}{i\sqrt{2}}[\hat{A}_{1}(t)-
\hat{A}^{\dagger}_{1}(t)], \label{sqz1}
\end{equation}
where
$\hat{A}_{1}(t)=\hat{a}_{1}(t)\exp[\frac{it}{2}(\omega_1+\omega_2)]$.
As we mentioned in section 1 we restrict the discussion to the first
mode.
The quadrature operators (\ref{sqz1}) satisfy the following commutation
relation

\begin{equation}
\left[ \hat{X},\hat{Y}\right] =i. \label{sqz2}
\end{equation}
Therefore, the uncertainty relation related to (\ref{sqz2}) is

\begin{equation}
\langle (\triangle \hat{X})^{2}\rangle \langle (\triangle \hat{Y}
)^{2}\rangle \geq \frac{1}{4}, \label{sqz3}
\end{equation}
where the variance $\langle (\triangle \hat{X})^{2}\rangle=\langle
\hat{X}^{2}\rangle -\langle \hat{X}\rangle ^{2}$ and similarly
form can be given for $\langle (\triangle \hat{Y})^{2}\rangle$.
We say that the system exhibits $X$-quadrature squeezing when

\begin{equation}
S=2\langle (\triangle \hat{X}(t))^2\rangle -1\leq 0. \label{sqz4}
\end{equation}
The equality sign holds, i.e. $S=0$, for minimum-uncertainty states.
 Similar definition can be given for the $Y$-quadrature
(defining a $Q$-factor).

On the other hand, the quadratures (\ref{sqz1}) can be represented
by the Hermitian operator

\begin{equation}
 \hat{V}_{\phi}=\frac{1}{\sqrt{2}}[\hat{A}_{1}(t)\exp(-i\phi)+
\hat{A}^{\dagger}_{1}(t)\exp(i\phi)], \label{frr1}
\end{equation}
where $\phi$ is a phase, which can be controlled by the homodyne
detector. The operator (\ref{frr1}) reduces to the $X$-quadrature
(the in-phase component of the field) when $\phi=0$, while when
$\phi=\pi/2$, it gives the $Y$-quadrature (out-of-phase component
of the field). According to (\ref{frr1}) the variance is dependent
on the phase $\phi$, which can be controlled to give the minimum
of all quadrature variances \cite{luks,tanas}. Thus we obtain the
notion of the principal squeezing \cite{luks,tanas}, which can be
expressed for the single-mode as
\begin{equation}
\eta(t)=2[ \langle \hat{A}_{1}^{\dagger}(t)
\hat{A}_{1}^{}(t)\rangle- \langle \hat{A}_{1}^{\dagger }(t)
\rangle \langle  \hat{A}_{1}(t)\rangle- |\langle
\hat{A}_{1}^{2}(t) \rangle- \langle \hat{A}_{1}(t) \rangle^{2}|].
\label{frr2}
\end{equation}
 Squeezing occurs when $\eta(t)<0$.

 The
different moments of the operator $\hat{A}_{1}(t)$ can be
evaluated as
\begin{equation}
 \langle \hat{A}_{1}^{\dagger m}(t) \hat{A}_{1}^{n}(t)\rangle
=\bar{\alpha}^{n}_{1}(t) \bar{\alpha}^{* m}_{1}(t)
z^{[\frac{n}{2}(n-1)-\frac{m}{2}(m-1)]}\exp[ \varepsilon
(z^{n-m}-1)],\label{f1}
\end{equation}
where
\begin{equation}
 \varepsilon=|\alpha_{1}|^{2}+|\alpha_{2}|^{2},\quad
z=\exp(-2i\chi t), \quad \bar{\alpha}_{1}(t)=
\bar{\alpha}_{x}(t)+i\bar{\alpha}_{y}(t)
 \label{f2}
\end{equation}
and
\begin{equation}
\bar{\alpha}_{x}(t)=
\alpha_{1}\cos (\lambda t), \quad
\bar{\alpha}_{y}(t)=
-[\alpha_{1}\frac{\Delta }{2}
+\alpha_2\kappa ]\frac{\sin (\lambda t)}{\lambda },
 \label{ff2}
\end{equation}
where $\alpha_1$ and $\alpha_2$ are the initial field amplitudes.
Throughout the paper we consider that $\alpha_1$ and $\alpha_2$
are real.
\begin{figure}
  \includegraphics[width=.8\linewidth]{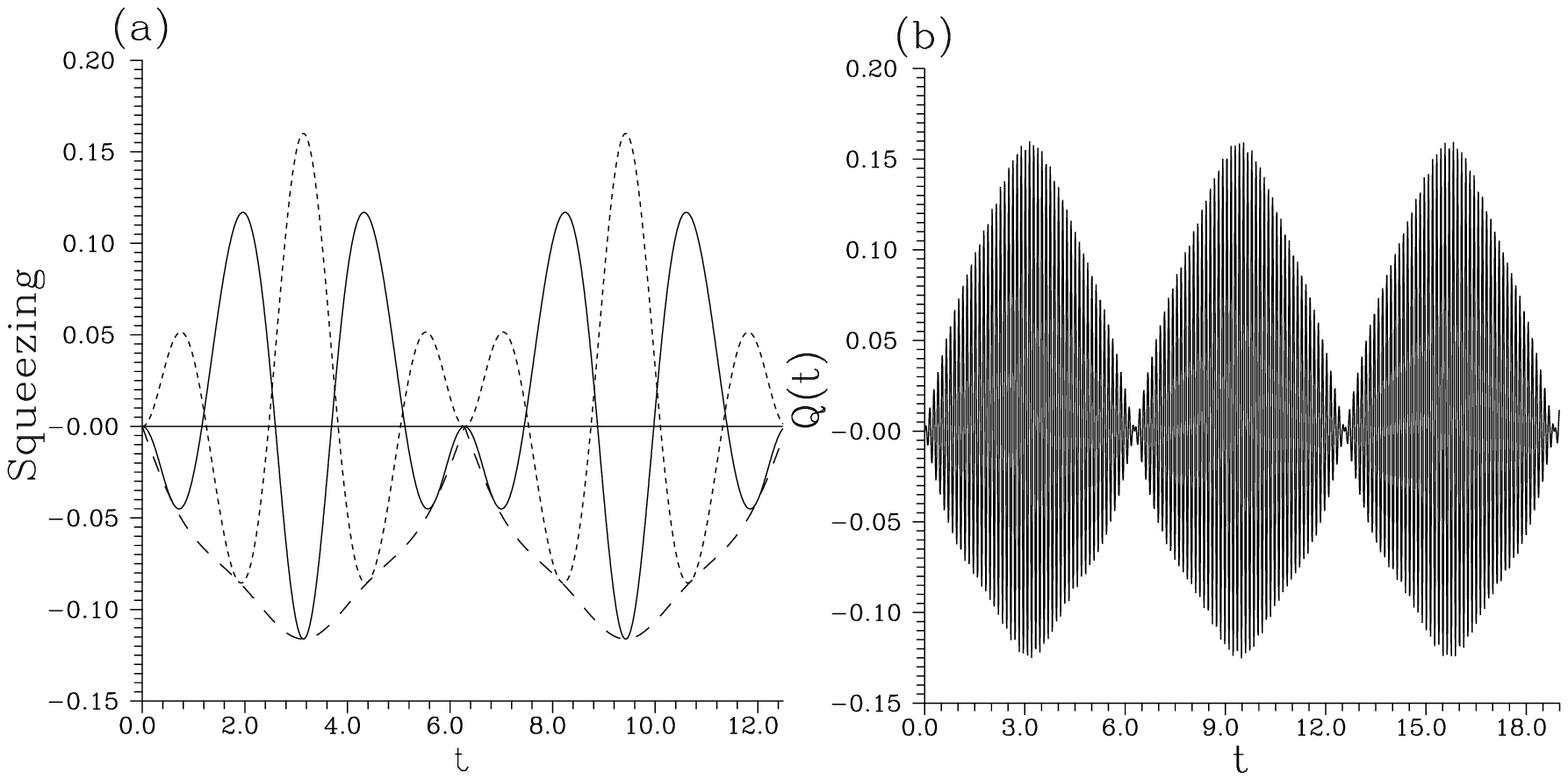}
 \caption{
Evolution of the squeezing factors  and principal squeezing of the
first mode when
 $\kappa=1, \chi =0.5, (\alpha_1,\alpha_2)=(0.2,0.2)$ and for
(a) $\Delta=0 s^{-1}$ (solid curve for $S(t)$, short-dashed curve
for $Q(t)$ and long-dashed curve for $\eta(t)$), (b) evolution of
the factor $Q(t)$ for $\Delta=50 s^{-1}$. The straight line in (a)
is given to show the squeezing bound.}
\end{figure}
Expression (\ref{f1}) reflects two facts: (i) When $\chi=0$
the system reduces to the up conversion process, which switches the
energy
between the waveguides without generating nonclassical effects.
 (ii) When $n=m$ all the moments are
independent of $\chi$ and the mean-photon numbers exhibit
oscillatory behaviour only rather than collapse-revival pattern,
which is representative for the DKNC \cite{hora2,
qu14,qu15,qu18,faisal1, ar1}. Moreover, this fact shows that the
system exhibits always Poissonian statistics.

Now we start the discussion by investigating the behaviour of the
quadratures squeezing. We show that the system can provide
squeezing and the quadrature squeezing can exhibit the
collapse-revival phenomenon. Furthermore, squeezing cannot be
simultaneously generated in the two quadratures. In doing so we
use (\ref{f1}) to express the squeezing factors $S(t)$ and $Q(t)$
as

\begin{eqnarray}
\begin{array}{lr} S(t)=
2|\bar{\alpha}_{1}(t)|^{2}+G_1(t)-G_2(t),
\\
\\
Q(t)= 2|\bar{\alpha}_{1}(t)|^{2}- G_1(t)- G_3(t), \label{f43}
\end{array}
\end{eqnarray}
where
\begin{eqnarray}
\begin{array}{lr} G_1(t)=2[(\bar{\alpha}^{2}_{x}(t)-
\bar{\alpha}^{2}_{y}(t))\cos(\epsilon(2\chi t))
+2\bar{\alpha}_{x}(t)
\bar{\alpha}_{y}(t)\sin(\epsilon(2\chi t))]f(2\chi t), \\
\\
G_2(t)=4[\bar{\alpha}_{x}(t)\cos(\epsilon(\chi t))
+\bar{\alpha}_{y}(t)\sin(\epsilon(\chi t))]^{2}
f^{2}(\chi t),
 \\
 \\
G_3(t)=4
[\bar{\alpha}_{x}(t)\sin(\epsilon(\chi t))
-\bar{\alpha}_{y}(t)\cos(\epsilon(\chi t))]^{2}
f^{2}(\chi t),
\\
\\
\epsilon(n\chi t))=n(n-1)\chi t+\varepsilon \sin(2n\chi t),\\
\\
f(n\chi t) =\exp[-2\varepsilon\sin ^{2}(n\chi t)]. \label{medaa}
\end{array}
\end{eqnarray}
It is obvious that the behaviours of $S(t)$ and $Q(t)$ are
periodic as a result of the nature of the coupler, which basically
depends on switching energy between waveguides by means of the
evanescent waves. Moreover, the origin of the occurrence of the
nonclassical effects is in the third order nonlinearity, which is
related to the envelope function $f(n\chi t)$, i.e. to the
nonlinear phase modulation term, that causes such effects.
Furthermore, involving squeezing factors, the function $f(n\chi
t)$ indicates that they can exhibit collapse-revival phenomenon,
as we shall see. Furthermore, the generation of squeezed light in
DKNC is quite obvious from (\ref{2}), which--apart from the free
part--includes two main parts: linear-interaction part and
self-cross nonlinear interaction part. The latter is well known in
the literature, see e.g. \cite{buz1}, as being able to generate
nonclassical light such as squeezed light, whereas the former
switches energy only. Now we discuss some analytical results for
the system by  focusing
 the attention on a simple  case, when $\Delta =0 s^{-1},
 \alpha_{1}=\alpha_{2}= \alpha$.
\begin{figure}
  \includegraphics[width=.8\linewidth]{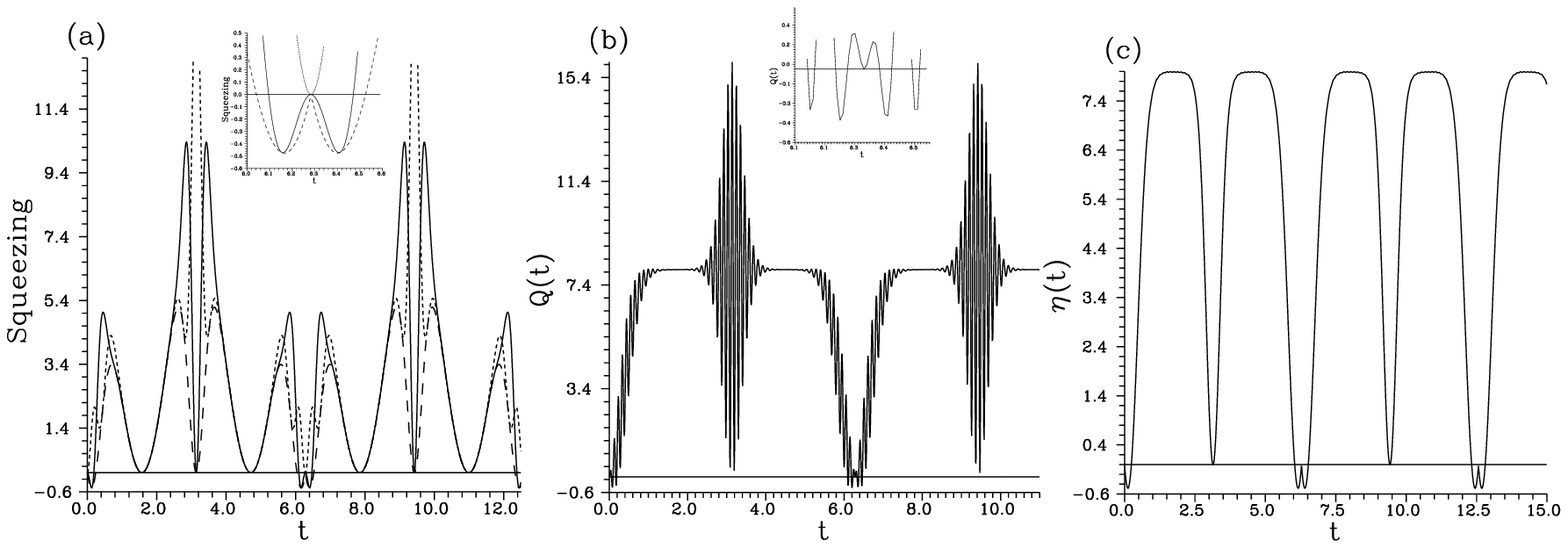}
 \caption{
Evolution of the squeezing factors and principal squeezing   of
the first mode when
 $\kappa=1, \chi =0.5, (\alpha_1,\alpha_2)=(2,0)$ and for
(a) $\Delta=0 s^{-1}$ (solid curve for $S(t)$, short-dashed curve
for $Q(t)$ and long-dashed curve for $\eta(t)$), and $\Delta=50
s^{-1}$ for $Q(t)$ (b) and $\eta(t)$ (c). The inset in (a) shows
that $Q(t)$ cannot exhibit squeezing, while that in (b) shows the
amount of squeezing, which can be obtained.}
\end{figure}
In this case expressions (\ref{f43}) can be written in the forms:
\begin{eqnarray}
\begin{array}{lr}
S(t)=2\alpha^{2}+2\alpha^{2}\cos[2t\lambda+\epsilon(2\chi t)]
f(2\chi t)-4\alpha^{2}\cos^{2}[t\lambda+\epsilon(\chi t)] f^{2}(\chi
t)],
\\
\\
Q(t)=2\alpha^{2}-2\alpha^{2}\cos[2t\lambda+\epsilon(2\chi t)]
f(2\chi t)-4\alpha^{2}\sin^{2}[t\lambda+\epsilon(\chi t)]
f^{2}(\chi t)].
\label{f44}
\end{array}
\end{eqnarray}
It is easy to prove that
\begin{equation}
S(t)+Q(t)=4\alpha^{2}[1-f^{2}(\chi t)].
\label{sqz5}
\end{equation}
It is evident that expressions (\ref{f44}) provide extreme values when
the envelope
functions tend to unity.
This means that when the system generates squeezed light
$S(t)+Q(t)\simeq 0$ and then squeezing cannot be
simultaneously generated in the both the quadratures.

The envelope function in the second term of (\ref{f44}) is maximum when
\begin{equation}
\chi t=m'\pi/2, \quad m'=1,2,...
\end{equation}
In this case expressions (\ref{f44}) reduce to
\begin{eqnarray}
\begin{array}{lr}
S(t)=
2\alpha^{2}\Bigl\{ [1-(-1)^{m'}]-2
[f^{2}(m'\frac{\pi}{2})-(-1)^{m'}]\cos^{2}(t\lambda)
\Bigr\},\\
\\
Q(t)=
2\alpha^{2}\Bigl\{ [1-(-1)^{m'}]-2
[f^{2}(m'\frac{\pi}{2})-(-1)^{m'}]\sin^{2}(t\lambda)
\Bigr\}.
\label{ff44}
\end{array}
\end{eqnarray}
\begin{figure}
  \includegraphics[width=.8\linewidth]{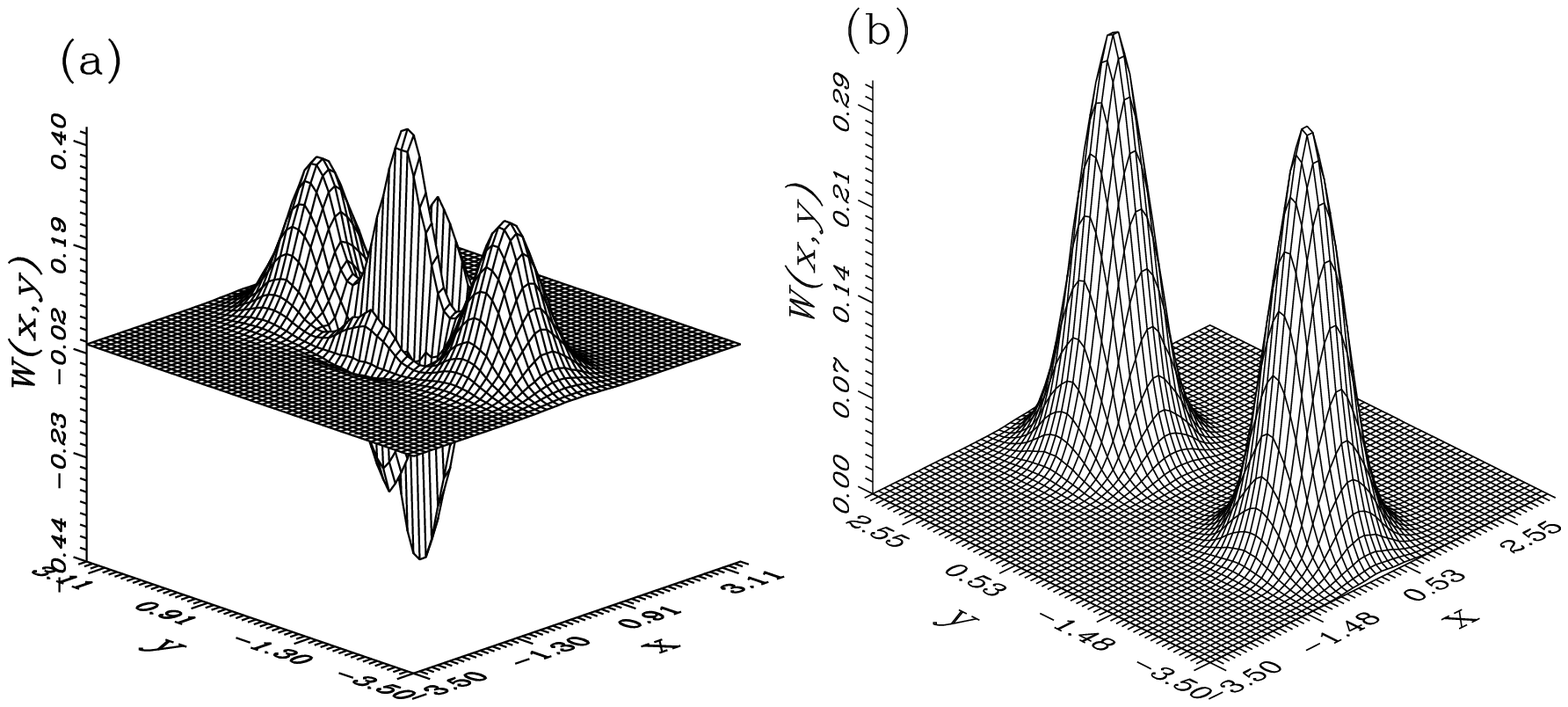}
 \caption{The
single mode $W$ function for $ \kappa =1, \chi =0.5,\Delta=0
s^{-1}, t=\pi $ and (a) $(\alpha_{1},\alpha_{2},D)=(2,0,0)$, (b)
$(2,2,4)$.}
\end{figure}
From (\ref{ff44}), for these particular values of the interaction
time, it is easy to prove that squeezing occurs for weak-intensity
regime, i.e. $0<\alpha<1$, where $f(m'\frac{\pi}{2})\simeq 1$ for
$m'=1,3,5,\cdots$. Nevertheless, squeezing cannot occur
for strong-intensity regime, i.e. when $\alpha>1$ where $f(\chi
t)\simeq 0$.
On the other hand, the third term in (\ref{f44}) is maximum when
\begin{equation}
\chi t=m'\pi, \quad m'=1,2,... \label{rab1}
\end{equation}
In this case (\ref{f44}) reduces to
$S(t)=Q(t)=0$, i.e. the system generate minimum-uncertainty states.
In spite of this fact we have numerically found that the system can
generate squeezing
close to $\chi t= m'\pi$ regardless
of the values of $\alpha_j$. This is related to the factor $\epsilon
(n\chi t)$ involving in the trigonometric functions.

\begin{figure}
   \includegraphics[width=.8\linewidth]{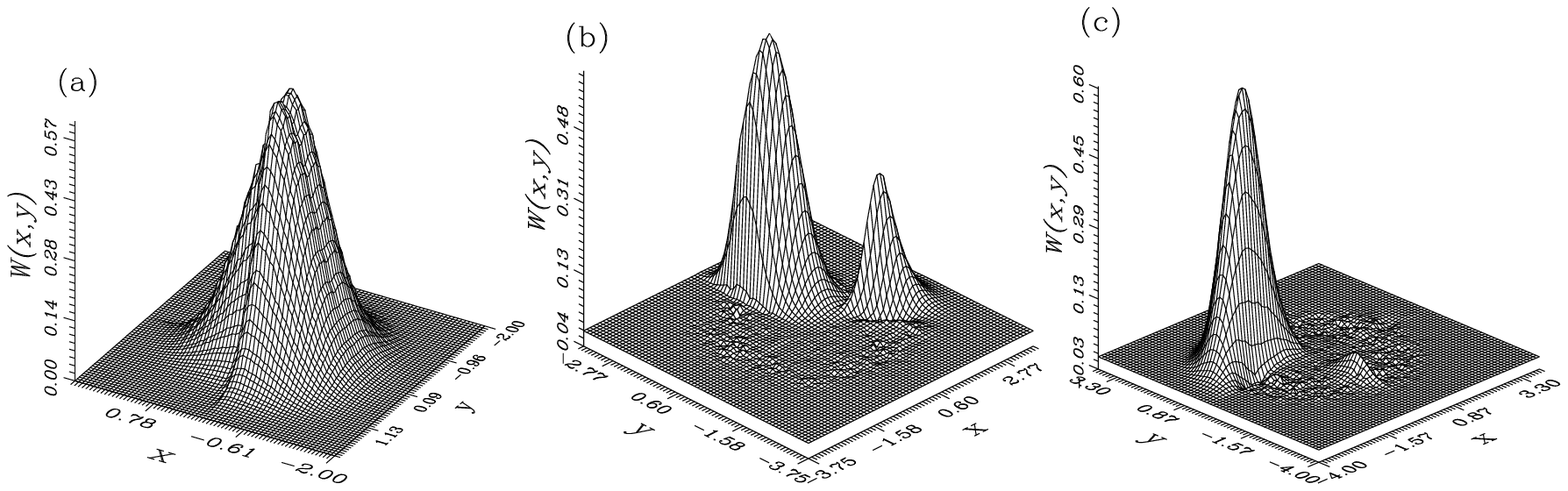}
 \caption{The
single mode $W$ function for $ \kappa =1, \chi =0.5$ and (a)
$(\Delta, \alpha_{1},\alpha_{2},t)=(0 s^{-1},0.2,0.2,3.139997)$,
(b) $(0 s^{-1},2,0,6.36005)$, (c) $(50 s^{-1},2,0,6.36005)$.}
\end{figure}

All these analytical facts and the influence of $\Delta$ on the
evolution of
quadrature squeezing
are presented in figures 1 and 2 for given values of
the parameters. From Fig. 1(a), where the intensities are weak,
one can see that squeezing periodically (but not simultaneously)
 occurs in both the quadratures.
The influence of the detuning parameter is given in Fig. 1(b),
which displays an exact periodic collapse-revival phenomenon. This
is related to the values of the scaled interaction time
$t\lambda$, which is rapidly changed when $\Delta$ is large. More
illustratively, the squeezing factors include two forms of
periodic functions: one is coming from the self-cross-nonlinear
interaction part, in particular, the envelope function whose
period is $\pi/\chi$, and the other is arising from the
linear-interaction part whose period is $\pi/\lambda$. Thus when
the values of $\Delta$ increase, the period of the energy exchange
between waveguides decreases, i.e. many oscillations occur, till
the interaction time becomes $t=\pi/\chi$ at this moment the field
is trapped instantaneously by nonlinearity in the waveguides and
the squeezing factors show collapse. As the interaction proceeds
the phenomenon is periodically repeated. Now we draw the attention
to Figs. 2. From Fig. 2(a) it is obvious that the system produces
vacuum and coherent light periodically. This can be realized
analytically since for $\Delta=0 s^{-1}, t=(m'+1/2)\pi, \lambda=1$
the amplitude $|\alpha_{1}(t)|= 0$ (regardless of the values of
the $\chi$) and the system produces the vacuum state.
 Similar arguments can be given for the coherent  light.
Also from Fig. 2(a) squeezing occurs in the $S(t)$ only (see the
solid curve in the inset) at particular values of the interaction
time. In fact, at these values the system generates Yurke-Stoler
coherent states (YSCS) \cite{yur},
 as we shall show in sections 3 and 4.
Moreover, squeezing can be established in $Q(t)$
 when the frequency mismatch is included (see Fig. 2(b)).
Also from Fig. 2(b) $Q(t)$ exhibits particular type of
collapse-revival phenomenon, which can be explained in the
following sense. According to the values of the interaction
parameters considered in Fig. 2(b) (i.e. $\alpha_2=0, \Delta\gg 1
s^{-1}$ and $\lambda\simeq \Delta/2$) expressions
(\ref{f44})--(\ref{rab1}) and the discussion around can be used.
In this regard $G_1(t)$ exhibits two times revival patterns
compared to those occurring in $G_3(t)$. When revivals occur in
$G_1(t)$ and $G_2(t)$ simultaneously, they destructively interfere
and cancel out each others showing such shape, however, at this
stage the system
 generates squeezing, i.e. squeezing occurs close to
$\chi =m'\pi$.
The comparison between figures 1(b) and 2(b) shows that as the values
of
$\alpha_j$ increase the widths (i.e. the envelopes of the revival
patterns)
decrease, but the collapse period is enlarged.
Actually, we found that $S(t)$ provides quite similar behaviour as
$Q(t)$.

Now we turn the attention to principal squeezing $\eta(t)$, which
for the mode under consideration can be expressed as

\begin{eqnarray}
\begin{array}{lr} \eta(t)=2|\bar{\alpha}_1(t)|^{2}\Bigl\{ 1-
\exp[-4\varepsilon \sin^{2}(\chi t)] -
\exp[-4\varepsilon \sin^{2}(\chi t)]\\
\\
\times \left[1+
 \exp[-8\varepsilon
\sin^{2}(\chi t)\cos(2\chi t)] -2 \exp[-4\varepsilon \sin^{2}(\chi
t)\cos(2\chi t)]
 \cos[2\chi t-4\varepsilon
\sin^{2}(\chi t)\sin(2\chi t)]\right]^{\frac{1}{2}}\Bigr\}.
 \label{frr3}
 \end{array}
\end{eqnarray}
As we did for the quadratures squeezing one can prove that for
$\chi t=m\pi$ the principal squeezing vanish, however, for $\chi
t=m'\pi/2$, $m'$ is odd integer, it reduces to

\begin{equation}
\eta(t)=-4|\bar{\alpha}_1(t)|^{2}\exp(-4\varepsilon). \label{frr5}
\end{equation}
This means that squeezing always occurs provided that $\alpha_j$
are finite. This is related to the fact that the phase of the
homodyne detector is adjusted for obtaining the minimum variance.
For the resonance case principal squeezing factor $\eta(t)$
 is presented by the long-dashed curves
 in Fig. 1(a) and Fig. 2(a)
 for weak and strong intensities, respectively. From Fig. 1(a) it is obvious that $\eta(t)$ provides
the envelope for the periodic nonclassical effects in the $Q(t)$
and $S(t)$ factors. More precisely, $\eta(t)$ starts from zero
before switching on the interaction, monotonically decreases
(increasing squeezing) as the interaction time increases till
$\chi t=\pi/2$ showing its minimum (maximum squeezing) and hence
increases monotonically till providing its initial value at $\chi
t=\pi$. This behaviour is periodically repeated. Similar behaviour
has been seen for the anharmonic oscillator model
\cite{luks,tanas}. From Fig. 2(a) for strong intensity regime
$\eta(t)$ exhibits behaviour, which is similar to that of $S(t)$.
Nevertheless, $\eta(t)$ provides nonclassical squeezing over
intervals of the interaction time larger than those for $S(t)$
(compare the long-dashed and solid curves in the inset).  Now we
draw the attention to the non-resonance case, which is presented
by Fig. 2(c) for given values of the interaction parameters. We
have noted that $\eta(t)$ cannot provide collapse-revival
phenomenon. Moreover, for weak intensity regime $\eta(t)$ is
insensitive to the value of
 $\Delta$, which  almost leads to the long-dashed curve in Fig. 1(a).
From Fig. 2(c), which is given for strong-intensity regime, we can
see that the behaviour is completely different from that of the
$Q(t)$, i.e. it is periodic  involving nonclassical effects
(squeezing) rather than collapse-revival phenomenon. Surprisingly,
Fig. 2(c) is similar to Fig. 2(e) \cite{qu14} for the resonance
case.

Generally, we can conclude that the origin of the occurrence
squeezing and collapse-revival phenomenon in the single mode
quadratures squeezing lies in the competition between the
nonlinearity and the frequency mismatch. The locations of the
revival patterns in the time domain depend on the values of
$t\chi$ and their shapes depend on the intensities of the field
launched in the waveguides initially. Finally, the behaviours of
the quadrature squeezing presented here are completely different
from those given in \cite{qu14}
 as a result of neglecting the
nonrelevant terms there. Also squeezing in the framework of the
principal squeezing has been remarked. Nevertheless, the principal
squeezing  cannot exhibit collapse-revival phenomenon.
\begin{figure}
  \includegraphics[width=.8\linewidth]{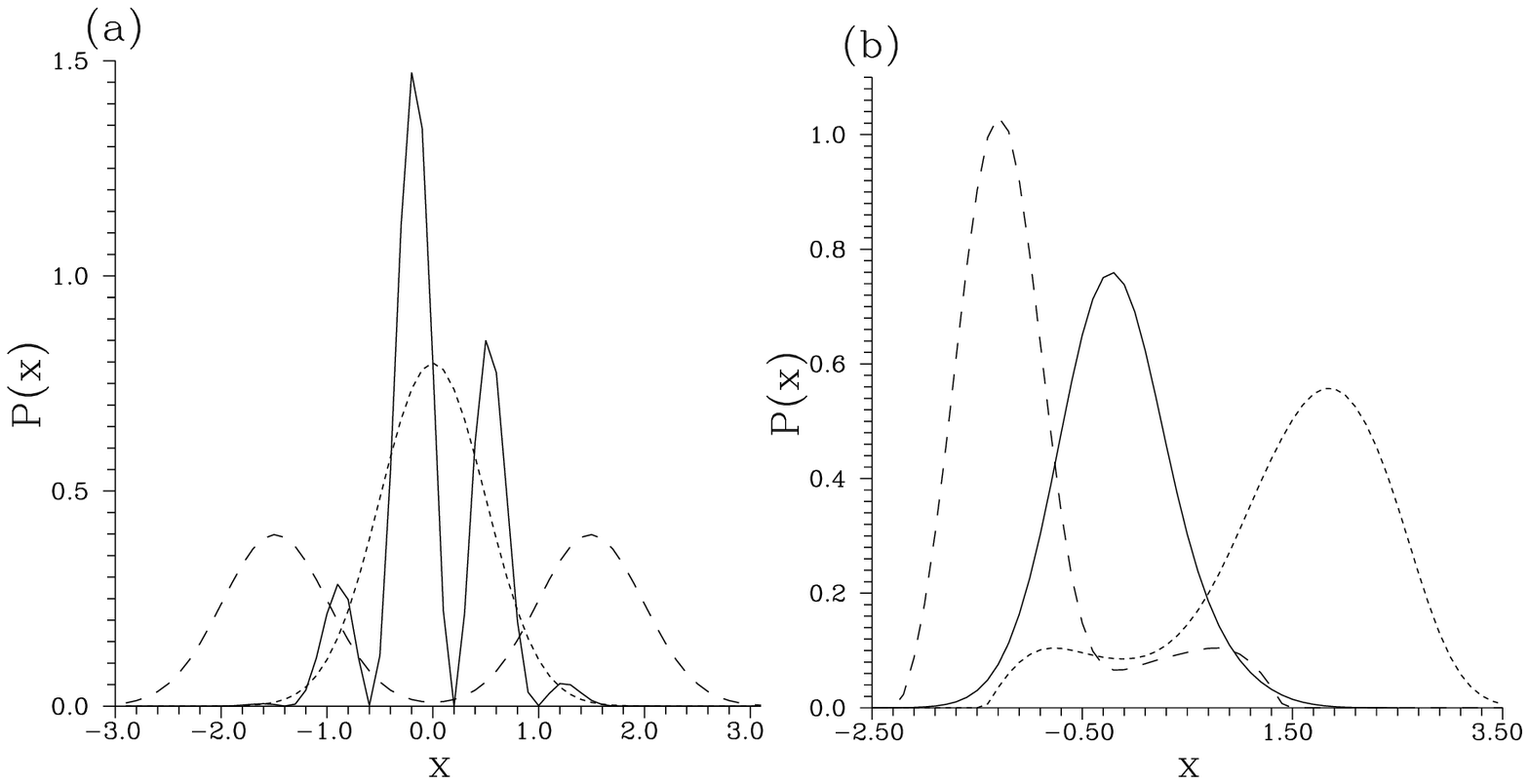}
  \caption{The
quadrature distribution $P(x)$ of the first mode for $\kappa=1,
\chi =0.5$ and for (a) $ t=\pi, \alpha_1=2$ and
$(\alpha_{2},\Delta^{2},D)=(0,0 s^{-2},0)$ (solid curve), $(2 ,0
s^{-2},4)$ (short-dashed curve) and $(0,5 s^{-2},0)$ (long-dashed
curve); (b) $(\Delta,\alpha_1,\alpha_2,t)=(0
s^{-1},0.2,0.2,3.139997)$ (solid curve), $(0 s^{-1},2,0,6.36005)$
(short-dashed curve) and $(50 s^{-1},2,0,6.36005)$ (long-dashed
curve).}
\end{figure}

\section{Quasiprobability distribution function}
Quasiprobability distribution functions ($W$-Wigner, $Q$-Husimi
and $P$-Glauber functions \cite{wign}) are important tools to give
insight in the statistical description of a quantum mechanical
system. These functions can be measured via homodyne tomography
\cite{tom}. Here we investigate the single mode quasiprobability
distribution functions and quadrature distribution for the system
under consideration using the technique given in \cite{El-O1}.
Furthermore, in the following section we use these functions in
investigating the phase distribution and phase variance. In order
to obtain these functions we have to calculate the single mode
$s$-parameterized characteristic function having the form
\begin{equation}
C(\zeta,t,s)={\rm Tr} \left\{ \hat{\rho}(0)\exp [\zeta
\hat{A}_{1}^{\dagger}(t)- \zeta^{*}
\hat{A}_{1}(t)+\frac{s}{2}|\zeta|^{2}] \right\}, \label{f65}
\end{equation}
where $\hat{\rho}(0)$ is the density
matrix of the system and
 $s$ is a parameter that takes on the values $1, 0$ and $-1$
corresponding
to normally, symmetrically and antinormally ordered characteristic
functions, respectively.
From (\ref{f1}) and (\ref{f65}) the characteristic function
 can be expressed as

\begin{eqnarray}
\begin{array}{lr}
C(\zeta,t,s)=\exp(\frac{s-1}{2}|\zeta|^{2})
\sum\limits_{n_{1},n_{2}=0}^{\infty}\frac{\zeta^{n_1}(-\zeta^{*})^{n_{2}}}
{n_1!n_2!}\bar{\alpha}^{n_2}_{1}(t) \bar{\alpha}^{*n_1}_{1}(t)\\
\\
\times z^{[\frac{n_2}{2}(n_2-1) -\frac{n_1}{2}(n_1-1)]}
\exp[\varepsilon (z^{n_2-n_1}-1)], \label{f6}
\end{array}
\end{eqnarray}
where we have considered that the two modes are initially prepared
in coherent states with amplitudes $\alpha_1, \alpha_2$.

The $s$-parameterized  quasiprobability functions can be obtained
through
the relation

\begin{equation}
W(\beta,t,s)=\frac{1}{\pi^{2}} \int
C(\zeta,t,s)
\exp(\beta\zeta^{*}- \beta^{*}\zeta)
d^{2}\zeta. \label{f7}
\end{equation}
On substituting (\ref{f6}) into (\ref{f7}) and using the technique
given in \cite{El-O1} we arrive at
\begin{eqnarray}
\begin{array}{lr}
W(\beta,t,s)=\frac{2}{\pi (1-s)}\exp(\frac{-2}{1-s}|\beta|^{2})
\sum\limits_{n_{1},n_{2}=0}^{\infty}\frac{(-1)^{n_1}\bar{\alpha}_{1}^{n_2}(t)
\bar{\alpha}_{1}^{* n_1}(t)}
{n_2!} \left(\frac{2}{1-s}\right)^{n_2}\beta^{* n_2-n_1}\\
\\
\times z^{[\frac{n_2}{2}(n_2-1) -\frac{n_1}{2}(n_1-1)]}
\exp[\varepsilon (z^{n_2-n_1}-1)] {\rm
L}_{n_1}^{n_2-n_1}\left(\frac{2|\beta|^{2}}{1-s}\right) ,
\label{f8}
\end{array}
\end{eqnarray}
where ${\rm L}^{m}_{n}(.)$ is the associated Laguerre polynomial of
order $n$
and $\beta=x+iy=|\beta|\exp(i\Theta)$, $\Theta$ being the phase
of $\beta$.
One can easily verify  that when $\chi t=m'\pi$ and $m'$ is integer,
the
 system produces coherent light with amplitude
 $\bar{\alpha}_{1}(t)$, as we mentioned in section 2.
This can be verified from (\ref{f8}) by means of the generating
function for the Leguerre polynomials (see (\ref{ap7}) in the
Appendix). In this case (\ref{f8}) reduces to the $W$ function for
the coherent state
\begin{equation}
W(\beta,t,s)=\frac{2}{\pi (1-s)}
\exp\left[ -\frac{2}{1-s}|\beta-\bar{\alpha}_{1}(t)|^{2}\right].
 \label{ff7}
\end{equation}
Furthermore, (\ref{ff7}) indicates that  the distribution cannot
provide negative values at the phase space origin, i.e. $\beta=0$.
Generally, (\ref{f8}) includes complicated quadratic phase factor,
which plays an essential role in generating cat states \cite{cat1}
and it reflects the strong entanglement between the two modes. In
fact, for particular values of the parameter $t\chi$ expression
(\ref{f8}) can reduce to that for the cat states. For instance,
when $t\chi=(m'+1/2)\pi$ and $m'$ is integer, (\ref{f8}) can be
modified to the following form:
\begin{eqnarray}
\begin{array}{lr} W(\beta,t,s)=\frac{1}{\pi (1-s)} \Bigl\{
\exp\left(-\frac{2}{1-s}|\beta-i\bar{\alpha}_1(t)|^{2}\right) +
\exp\left(-\frac{2}{1-s}|\beta+i\bar{\alpha}_1(t)|^{2}\right)\\
\\
+2\exp\left[-\frac{2}{1-s}(|\beta|^{2}+D
)\right]
\sin\left(\frac{2}{1-s}(\beta\bar{\alpha}_1^{*}(t)
+\beta^{*}\bar{\alpha}_1(t))\right)\Bigr\}, \label{im1}
\end{array}
\end{eqnarray}
where
\begin{equation}
D=\varepsilon -|\bar{\alpha}_1(t)|^{2}. \label{im2}
\end{equation}
The derivation of (\ref{im1}) is given in the Appendix.
Moreover, one can easily verify that
(\ref{im1}) is normalized,
\begin{equation}
\int W(\beta,t,s) d^{2}\beta =1. \label{imm1}
\end{equation}

\begin{figure}
  \vspace{0cm}
\centerline{\epsfxsize=8cm \epsfbox{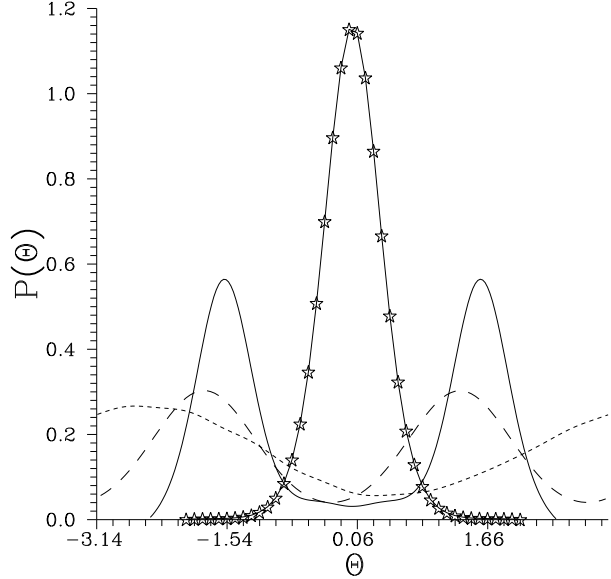} } \vspace{0.2cm}
\caption{The single mode phase distribution $P(\Theta,t,s=-1)$ for
$\kappa =1, \chi =0.5, (\Delta, \alpha_{1},\alpha_{2})=(0
s^{-1},2,2)$ and for $t=0$ (star-centered curve), $\pi/4$
(short-dashed curve), $2.94$ (long-dashed curve) and $\pi$ (solid
curve).}
\end{figure}

The form (\ref{im1}) reduces to that of YSCS \cite{yur} when
$D\simeq 0$. In other words, when $D\simeq 0$ the system generates
the states
\begin{equation}
|\psi(t)\rangle=\frac{1}{\sqrt{2}}[|i\bar{\alpha}_{1}(t)\rangle+\exp(i\frac{\pi}{2})
|-i\bar{\alpha}_{1}(t)\rangle], \label{imm2}
\end{equation}
which is YSCS. As is well known YSCS exhibit Poissonian statistics
and provide squeezing \cite{fai}. These results are obtained in
section 2. This indicates that the cat states generated in DKNC
are YSCS. Moreover, when $\Delta=0 s^{-1}$ and from (\ref{ff2}),
expression (\ref{im2}) takes the form
\begin{equation}
D=\alpha_1^{2}\sin^{2}(\lambda t)
+\alpha_2^{2}\cos^{2}(\lambda t). \label{imm4}
\end{equation}
Expression (\ref{imm4}) provides several consequences: (i) DKNC
generates YSCS
when one of the modes is initially prepared in the vacuum state and the
other is in the coherent state provided that the length of
the waveguides (or the interaction time) is controlled so that $D=0$.
(ii) when $\lambda=0$, i.e. the linear interaction between the
waveguides is neglected, the system can generate YSCS by choosing
$\alpha_1\neq 0$ and $\alpha_2=0$. (iii) the system cannot generate
YSCS when $\alpha_1=\alpha_2=0$ even though $D=0$, because the
expression
(\ref{im1}) reduces to that for the vacuum.

On the other hand, when $D\geq 1$ the interference part in (\ref{im1})
is
destroyed and hence the DKNC generates the statistical-mixture coherent
states
having the form
\begin{equation}
\hat{\rho}(t)=\frac{1}{2}\left[ |i\bar{\alpha}_{1}(t)\rangle
\langle i\bar{\alpha}_{1}(t)| +|-i\bar{\alpha}_{1}(t)\rangle
\langle -i\bar{\alpha}_{1}(t)|\right]. \label{imm3}
\end{equation}
An illustration of (\ref{im1}) (or (\ref{f8})) is given in Figs.
3(a) and (b) for $D=0$ and $4$, respectively. In Fig. 3(a) the $W$
function consists of two Gaussian bells, corresponding to the
statistical mixture of individual composite states (cf.
(\ref{imm2})) and interference fringes inbetween arising as a
result of the contribution originating from the quantum
superposition. In Fig. 3(b) the interference fringes are
completely suppressed and the form of the statistical-mixture
coherent states is well pronounced. Actually, by controlling the
values of the $\Delta$, the form of the cat states can be less or
more pronounced. For instance, Fig. 3(a) can be reduced to Fig.
3(b) by taking $\Delta=\sqrt{5} s^{-1}$ instead of zero.

Now we shed the light on the general case, e.g. apart from the
case $t\chi=(m'+1/2)\pi$. This will be numerically done by figures
4 for the $W$ function obtained from (\ref{f8}). The values of the
interaction parameters in Figs. 4 have been selected such that the
system provides quadratures squeezing (see Figs. 1 and 2). In all
these figures we can observe the nonclassical effects, such as
negative values, stretching, multipeak structure and deformation.
In Fig. 4(a), where the initial intensities of the modes are weak,
the $W$ function exhibits contour-stretching as well as
deformation around the phase space origin.  The origin of such
behaviour is in the generation of cat states in the microscopic
regime \cite{El-O1}, where the contribution of the different
components of the cat are located close to the phase space origin
competing each others. When the values of the intensities and the
value of the interaction time are increased, the shape of the cat
states becomes more pronounced by involving
 multipeak structure (asymmetric peaks) (see Fig. 4(b)).
This is in a good agreement with the information given above even
though
$t\chi\neq (m'+1/2)\pi$.
 We proceed by drawing the attention to the influence of
the detuning parameter on the behaviour of the $W$ function, which
is given in Fig. 4(c). In this figure the $W$ function exhibits
almost stretched single peak structure as well as negative values
indicating that the nonclassical effects are still occurring. The
transition from Fig. 4(b) to Fig. 4(c) through changing the values
of the detuning is related to the fact that in the former the
amplitude $\bar{\alpha}_1(t)$ is real, while in the latter it is
complex (one can check this for the chosen values of the
interaction parameters). This leads to that the oscillations
arising from the nonlinear phase modulation are superimposed with
those of the field amplitude $\bar{\alpha}_1(t)$ causing such
behaviour in Fig. 4(c). We conclude this part by mentioning that a
connection between the Fourier coefficients and the $Q$ function
 for the case $2\chi=\bar{\chi}$ using  the
diagonalization approach is given in \cite{qu18}.

We close this section by investigating the quadrature (or
position) distribution $P(x,t)$, which can be measured in the
homodyne detector \cite{yur}. The distribution $P(x,t)$ can be
evaluated via the $W$ function through the relation
\begin{equation}
P(x,t)=\int\limits_{-\infty}^{\infty} W(x+iy,t,0) dy. \label{qdx1}
\end{equation}
Substituting (\ref{f8}) into (\ref{qdx1}) and after lengthy
calculation, we arrive at
\begin{eqnarray}
\begin{array}{lr} P(x,t) =\sqrt{\frac{2}{\pi}}\exp(-2x^2)
\sum\limits_{n_{1},n_{2}=0}^{\infty}
\sum\limits_{r=0}^{min(n_1,n_2)}
\frac{(-2)^{r}2^{-\frac{1}{2}(n_1+n_2)}\bar{\alpha}_{1}^{n_2}(t)
\bar{\alpha}_{1}^{* n_1}(t)}
{r! (n_1-r)!(n_2-r)!} \\
\\
\times z^{[\frac{n_2}{2}(n_2-1) -\frac{n_1}{2}(n_1-1)]} \exp[
\varepsilon (z^{n_2-n_1}-1)] {\rm H}_{n_1-r}(\sqrt{2}x) {\rm
H}_{n_2-r}(\sqrt{2}x) , \label{qxd2}
\end{array}
\end{eqnarray}
where ${\rm H}_{m}(.)$ is the Hermite polynomial of order $m$. It
is more convenient to give the explicit form for $P(x,t)$ for the
case $t\chi=(m'+1/2)\pi$
\begin{eqnarray}
\begin{array}{lr}
P(x,t)
=\frac{1}{\sqrt{2\pi}}\Bigl\{\exp[-2(x+\bar{\alpha}_y(t))^2]
+\exp[-2(x-\bar{\alpha}_y(t))^2]\\
\\
+2\exp[-2(x^2+\bar{\alpha}_y^2(t)+D)]\sin(4x\bar{\alpha}_x(t))\Bigr\}.
\label{qxd3}
\end{array}
\end{eqnarray}
In Figs. 5(a) and (b) we have plotted $P(x,t)$ corresponding to
the $W$ functions shown in Figs. 3 and 4, respectively. In general
one can observe that $P(x,t)$ provides nonclassical effects by
including oscillatory behaviour. In Fig. 5(a) the solid curve
shows the oscillatory behaviour related to YSCS, which is a direct
consequence of the interference in phase space. Nevertheless, the
short-dashed curve is a Gaussian bell having its maximum value at
$x=0$. This indicates that the system generates vacuum light.
Actually, for this curve $\bar{\alpha}_y(t)\simeq 0$ and the
interference part is completely suppressed since $D=4$. This may
contradict to the result illustrated in Fig. 3(b) for the
corresponding $W$ function, in which the generation of the
statistical-mixture coherent state is obvious. This confusion can
be removed if the attention is drawn to the momentum distribution
$P(y,t)$, which exhibits, for this case, a two-peak structure (we
have checked this fact). The influence of the $\Delta$ is
presented by the long-dashed curve, which provides the two-peak
structure indicating the generation of the statistical-mixture
coherent state. On the other hand, we can note that there is
almost  consistence between the behaviours in Figs. 4 and Fig. 5b,
where $P(x,t)$ provides two asymmetric peaks for strong
intensities regardless of the values of $\Delta$ and a single peak
for weak intensities. We conclude this section by mentioning in
\cite{sch} stated that "the concept of interference in phase space
 readily explains the similarity between the photon number distribution
and the position distribution".  This is not always correct. For
instance, $P(x,t)$ of the YSCS provides oscillatory behaviour (see
the solid curve in Fig. 5(a)) even though the photon number
distribution of YSCS is Poissonian, as we mentioned in section 2.

\section{Phase distribution and its variance}
In this section we study the evolution of the phase distribution
and its variance. The phase distribution associated with the first
mode can be obtained from the quasidistribution (\ref{f8}) by
integrating $W(\beta,t,s)$ over the radial variable as
\begin{equation}
P(\Theta,t,s)=\int_{0}^{\infty}|\beta|W(\beta,t,s)d|\beta|. \label{f9}
\end{equation}
On substituting (\ref{f8}) into (\ref{f9}) and carrying out the
integration,
we arrive at
\begin{eqnarray}
\begin{array}{lr} P(\Theta,t,s)=\frac{1}{2\pi}\Bigl\{ 1+
2\sum\limits_{n_{2}>n_{1}}^{\infty}\sum\limits_{m=0}^{n_1}
\frac{(-1)^{n_{1}+m}\Gamma(m+\frac{n_2-n_1}{2}+1)
|\bar{\alpha}_{1}(t)|^{n_2+n_1}} {(n_1-m)!(n_2-n_1+m)!m!}
\left(\frac{2}{1-s}\right)^{\frac{n_1+n_2}{2}}
\\
\\
\times \cos[\varphi +(n_1-n_2)\Theta] \exp[-2\varepsilon
\sin^2((n_2-n_1)\chi t)]\Bigr\}, \label{f11}
\end{array}
\end{eqnarray}
where
\begin{eqnarray}
\begin{array}{lr} \bar{\alpha}_{1}(t)=
|\bar{\alpha}_{1}(t)|\exp[i\bar{\phi}(t)], \\
\\
\varphi=(n_2-n_1)\bar{\phi}(t)+[n_2(n_2-1)-n_1(n_1-1)]\chi
t+\varepsilon \sin(2\chi t(n_2-n_1)) \label{ff11}
\end{array}
\end{eqnarray}
and $\Gamma (.)$ is the Gamma function. The phase distribution
given by (\ref{f11}) is just a special case of the formula (2.59)
in \cite{tanas2}.
\begin{figure}
   \includegraphics[width=.8\linewidth]{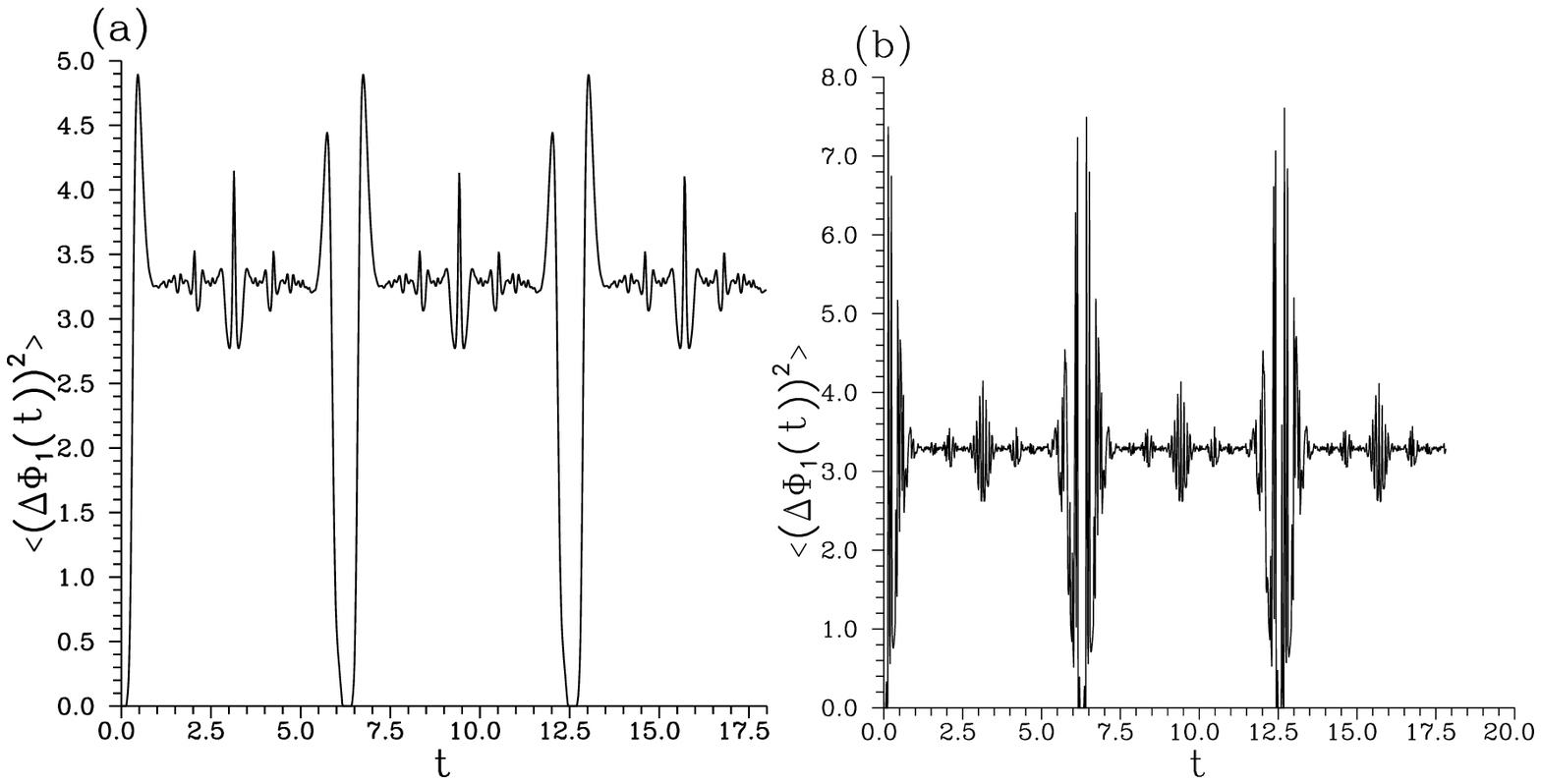}
  \caption{ The
single mode phase variance $\langle (\triangle
\Phi_{1}(t))^{2}\rangle$  against the interaction time $t$ for
$\kappa =1, \chi =0.5$ and (a) $(\Delta, \alpha_{1},\alpha_{2})=(0
s^{-1},2,2)$ and (b) $(50 s^{-1},2,2)$ .}
\end{figure}
As is well known the phase distribution obtained from the
quasiprobability distribution functions includes some
difficulties. For instance, the singularity of the $P$ function
and the negative values involved in the $W$ function for some
quantum mechanical systems may be reflected in the corresponding
phase distributions (see, e.g., \cite{scr}). These difficulties
can be avoided by using $Q$ function, which is well defined and is
always positive, in evaluating the phase distribution.
Illustration of $P(\Theta,t,s=-1)$ is given in Fig. 6 for given
values of the parameters. From this figure one can see how the cat
states can be generated in the system depending on the values of
the interaction time (or the length of the waveguides). For
instance, at $t=0$, i.e. before switching on the interaction,
$P(\Theta)$ exhibits single-peak structure around $\Theta=0$,
which is representative for the coherent light (see the
star-centered curve). As the interaction proceeds, i.e. energy
exchange between the waveguides starts to play a role, the initial
peak reduces to two wings around $\Theta=\pm \pi$, which becomes
straight line at $t\chi=\pi/2$, i.e. the phase distribution
exhibits normal distribution (this case has not been included in
the figure). It is worth remembering that the system generates
vacuum state in this case. When the interaction time becomes close
to $\pi$ the form of the cat states starts to appear as a broader
two-peak structure (see the long-dashed curve in Fig. 6).
Eventually, at $t=\pi$ the distribution provides two Gaussian
peaks
 around $\Theta=\pm\pi/2$ indicating the generation of
the cat state, in particular, the statistical-mixture coherent
state (cf. (\ref{imm3})). Actually, the phase distribution is
insensitive to the interference in phase space, i.e. the phase
distribution of the even, odd, Yurke-Stoler and the
statistical-mixture coherent states are almost similar
\cite{buz2}. This leads to the fact that the phase distribution is
insensitive to the values of $\Delta$, which can change cat states
to the statistical-mixture coherent states (see the discussion
given in section 3 for the $W$ function).

The single mode phase variance is defined as
\begin{equation}
\langle(\triangle\hat{\Phi}_{1}(t))^{2}\rangle=
\langle\hat{\Phi}_{1}^{2}(t)\rangle-\langle
\hat{\Phi}_{1}(t)\rangle^{2}.\label{15b}
\end{equation}
The $l$th moment of the phase distribution can be evaluated from
(\ref{f11}) through the relation
\begin{equation}
\langle\hat{\Phi}^{l}_{1}(t)\rangle=
\int\limits_{-\pi}^{\pi}\Theta^l P(\Theta,t,s) d\Theta.\label{15bb}
\end{equation}
The evolution of the
$\langle(\triangle\hat{\Phi}_{1}(t))^{2}\rangle$ is given in Figs.
7(a) and (b) for $\Delta=0 s^{-1}$ and $50 s^{-1}$, respectively.
In Fig. 7(a) the periodic behaviour is dominant indicating
switching of energy between the waveguides. Also one can see that
the system provides its initial stage periodically. The detuning
parameter $\Delta$ reflects itself in
$\langle(\triangle\hat{\Phi}_{1}(t))^{2}\rangle$ as
collapse-revival-subrevival phenomenon (see Fig. 7(b)). As we
mentioned in section 2, when the values of the $\Delta$ increase,
the period of switching of energy between the waveguides
decreases, causing such phenomenon. Furthermore, the main revivals
occur around the values of the interaction times at which the
system reduces to its initial form, however, the subrevivals occur
when the system generate cat states. Actually, the density matrix
of the superposition states has different components, each of them
has its own collapse-revival pattern (when $\Delta$ is large),
which interfere with each others producing subrevivals. Thus we
can conclude that the occurrence of the
collapse-revival-subrevival phenomenon is a direct consequence of
the generation of different types of states in the system,
 i.e. vacuum, coherent, and cat states.
It is worth mentioning that such behaviour has been
observed for the single mode phase variance of the two-mode
Jaynes-Cummings
model \cite{comin}, however, the behaviour presented here is more
systemic.

\section{Conclusion}
In this paper we have discussed the single mode quantum properties for
the codirectional nonlinear Kerr coupler, when the frequency
mismatch is involved.
The attention is focused on the case $2\chi=\widetilde{\chi}$ for which
the solutions of the equations of motion are exact.
 We have investigated quadratures squeezing, principal
squeezing, quasiprobability distribution functions, quadrature
distribution, phase distribution and phase variances. Generally,
we have shown that the light obtained from the system exhibits
Poissonian statistics and provides squeezing in the framework of
quadratures and principal squeezing. Moreover, it has been shown
that when the values of $\Delta$ increase the period of the energy
exchange between waveguides decreases. This fact leads to many
interesting effects, such as the quadratures squeezing and the
phase variances can exhibit collapse-revival and
collapse-revival-subrevival phenomena,
 respectively.
The generation of YSCS in the system has been analytically and
numerically
demonstrated and confirmed in all studied quantities.
 YSCS can be generated even if the linear interaction
between the waveguides is neglected. Moreover, we have shown that
the generation of these states most probably occurs when initially
one of the modes is in the coherent state while the other is in
vacuum state, provided that $t\chi =N\pi$ and $N$ is a fraction of
integer. Furthermore, the system can generate the
statistical-mixture coherent states in dependence on the values of
the interaction parameters. The nonclassical effects have been
remarked in the behaviour of $P(x,t)$. Finally, $P(x,t)$ cannot
include a complete information on the interference in phase space.

\section*{\bf Appendix}

In this appendix we give the derivation of the quasiprobability
distribution given by (\ref{im1}).
It is worth reminding that $t\chi=(m'+1/2)\pi$, i.e. $z=-1$.
We start by providing the following array:
\begin{eqnarray}
n_1&=& 1,2,3,4,5,6,7,8,\cdots\nonumber\\
l=\frac{n_1}{2}(n_1-1)&=& 0,0,1,3,6,10,15,21,\cdots\nonumber \\
z^l&=& 1,1,-1,-1,1,1,-1,-1,\cdots \label{ap1}
\end{eqnarray}
From the information shown in (\ref{ap1})
we can express the summation associated with the index $n_1$ in
(\ref{f8}) (for this case) as
\begin{equation}
W(\beta,t\chi,s)=\sum\limits_{n_1,n_2}(-1)^{n_1}
z^{-\frac{n_2}{2}(n_2-1)} [h(2n_1-n_2) F(2n_1,n_2)-h(2n_1+1-n_2)
F(2n_1+1,n_2)], \label{ap2}
\end{equation}
where

\begin{equation}
F(n_1,n_2)=\frac{2}{\pi
(1-s)}\exp\left(\frac{-2}{1-s}|\beta|^{2}\right)
\frac{\bar{\alpha}_{1}^{n_2}(t) \bar{\alpha}_{1}^{* n_1}(t)}
{n_2!} \left(\frac{2}{1-s}\right)^{n_2}\beta^{* n_2-n_1} {\rm
L}_{n_1}^{n_2-n_1}\left(\frac{2|\beta|^{2}}{1-s}\right),\label{ap3}
\end{equation}
\begin{equation}
h(n_1-n_2)=\exp[\varepsilon (z^{n_2-n_1}-1)].\label{ap4}
\end{equation}
It is evident that, for these specified values of $t\chi$, when
$n_1-n_2$ is even $h(n_1-n_2)=1$ otherwise
$h(n_1-n_2)=\exp(-2\varepsilon )$. Moreover, the summation related
to $n_2$ in (\ref{ap2}) can be similarly expressed as that over
$n_1$ and hence (\ref{ap2}) takes the form
\begin{eqnarray}
\begin{array}{lr}
W(\beta,t\chi,s)=\sum\limits_{n_1,n_2}(-1)^{(n_1+n_2)}\\
\Bigl\{F(2n_1,2n_2)-F(2n_1+1,2n_2+1)\\
\\
+ \exp[-2\varepsilon ]
[F(2n_1,2n_2+1)-F(2n_1+1,2n_2)]\Bigr\}.\label{ap5}
\end{array}
\end{eqnarray}
Now we show how the first summation can be evaluated in a closed
form:

\begin{equation}
\sum\limits_{n_1,n_2}(-1)^{(n_1+n_2)} F(2n_1,2n_2)
=\frac{1}{4}\sum\limits_{n_1,n_2}(i)^{(n_1+n_2)}
[1+(-1)^{n_1}][1+(-1)^{n_2}]F(n_1,n_2). \label{ap6}
\end{equation}
Substitute (\ref{ap3}) into (\ref{ap6}) and use the generating
function for the Laguerre polynomials \cite{tab} as

\begin{equation}
\exp(-xk)(1+k)^{\nu}=
\sum\limits_{n}k^{n} {\rm L}_n^{\nu-n}(x).  \label{ap7}
\end{equation}
Therefore, the summation related to the index $n_1$ in (\ref{ap6})
can be easily carried out and we obtain
\begin{eqnarray}
\begin{array}{lr} \sum\limits_{n_1,n_2}(-1)^{(n_1+n_2)}
F(2n_1,2n_2) =\frac{1}{2\pi
(1-s)}\sum\limits_{n_2}\frac{\left[\frac{2i\bar{\alpha}_1(t)\beta^{*}}{(1-s)}\right]^{n_2}}
{n_2!} [1+(-1)^{n_2}]
\Bigl\{ [1+\frac{i\bar{\alpha}_1^{*}(t)}{\beta^{*}}]^{n_2}\\
\\
\times \exp\left(-\frac{2i\bar{\alpha}_1^{*}(t)\beta}{1-s}\right)
+ [1-\frac{i\bar{\alpha}_1^{*}(t)}{\beta^{*}}]^{n_2}

\exp\left(\frac{2i\bar{\alpha}_1^{*}(t)\beta}{1-s}\right)\Bigr\}.\label{ap8}
\end{array}
\end{eqnarray}
Now the summation over $n_2$ can be straightforwardly evaluated
and we obtain
the following closed form expression:
\begin{eqnarray}
\begin{array}{lr} \sum\limits_{n_1,n_2}(-1)^{(n_1+n_2)}
F(2n_1,2n_2) =\frac{1}{2\pi (1-s)} \Bigl\{
\exp\left(-\frac{2}{1-s}|\beta-i\bar{\alpha}_1(t)|^{2}\right) +
\exp\left(-\frac{2}{1-s}|\beta+i\bar{\alpha}_1(t)|^{2}\right)\\
\\
+2\exp\left[-\frac{2}{1-s}(|\beta|^{2}-|\bar{\alpha}_1(t)|^{2})\right]
\cos\left(\frac{2}{1-s}(\beta\bar{\alpha}_1^{*}(t)
+\beta^{*}\bar{\alpha}_1(t))\right)\Bigr\}. \label{ap9}
\end{array}
\end{eqnarray}
Similar procedures have to be performed to obtain the other terms
in (\ref{ap5}) and then (\ref{im1}) is obtained.

\section*{Acknowledgement}
J.P. thanks the partial support from the grant LN00A015 of the
Czech Ministry of Education and from the EU Project COST OCP
11.003.

\section*{References}

\end{document}